\documentclass[12pt]{article}
\usepackage{epsfig}
\begin{document}

\baselineskip=.22in
\renewcommand{\baselinestretch}{1.2}
\renewcommand{\theequation}{\thesection.\arabic{equation}}

\begin{flushright}
{\tt hep-th/0301076}
\end{flushright}

\vspace{5mm}

\begin{center}
{{\Large \bf Electromagnetic String Fluid in Rolling Tachyon}\\[12mm]
{Chanju Kim\footnote{cjkim$@$ewha.ac.kr}, 
Hang Bae Kim\footnote{HangBae.Kim$@$ipt.unil.ch}, 
Yoonbai Kim\footnote{yoonbai$@$skku.ac.kr}, and
O-Kab Kwon\footnote{okwon$@$newton.skku.ac.kr}}\\[7mm]
{\it ${}^{1}$Department of Physics, Ewha Womans University,
Seoul 120-750, Korea}\\[2mm]
{\it ${}^{2}$Institute of Theoretical Physics, University of Lausanne,
CH-1015 Lausanne, Switzerland}\\[2mm]
{\it ${}^{3,4}$BK21 Physics Research Division and Institute of Basic Science,
Sungkyunkwan University,\\
Suwon 440-746, Korea}\\[2mm]
{\it ${}^{3}$ School of Physics, Korea Institute for Advanced Study,\\
207-43, Cheongryangri-Dong, Dongdaemun-Gu, Seoul 130-012, Korea}
}
\end{center}

\vspace{5mm}

\begin{abstract}
We study Born-Infeld type effective action for unstable D3-brane system
including a tachyon and an Abelian gauge field, and find the rolling tachyon
with constant electric and magnetic fields as the most general homogeneous 
solution. Tachyonic vacua are characterized by magnitudes of the electric 
and magnetic fields and the angle between them. Analysis of small 
fluctuations in this background shows that the obtained configuration may be
interpreted as a fluid consisting of string-like objects carrying electric
and magnetic fields. They are stretched along one direction and 
the rolling tachyon move in a perpendicular direction to the strings. 
Direction of the propagating waves coincides with that of strings with 
velocity equal to electric field.
\end{abstract}


\newpage

\setcounter{equation}{0}
\section{Introduction}

Instability of an unstable D-brane or a pair of D-brane-anti-D-brane
is manifested by a tachyonic mode and, in the context of effective field 
theory, decay of the unstable brane is depicted by condensation of  
tachyon~\cite{oSen}.
As the tachyon approaches its true vacuum where brane tension is dissipated
away, open string degrees of freedom disappear and correspondingly 
perturbative plane-wave type excitations are completely absent in the effective
field theory. 
The classical decay process is described by the so-called rolling tachyons
which are constructed as a family of spatially-homogeneous but time-dependent 
classical solutions in open string theory~\cite{Sen,MZ}.

This rolling tachyon has been mainly applied to various topics in
cosmology in the very early universe, e.g., inflation, dark matter, and
reheating~\cite{Gib,KL,tacos1,KKK}. In tachyon cosmology or other applications,
inclusion of bosonic string degrees looks indispensable, e.g., gauge field, 
graviton, dilaton, antisymmetric tensor field, or RR-field. In relation 
with string dynamics, U(1) gauge field on the brane has played an  
important role~\cite{Yi}, of which
dynamics is given by Born-Infeld type 
gauge kinetic term in the effective action on the 
brane~\cite{Gar,KMM} since one of two U(1) gauge symmetries remains
unbroken in the decaying D$\bar{{\rm D}}$-system despite of tachyon
condensation~\cite{Wit}. In pure rolling tachyon background, perturbative
electromagnetic waves stop propagating around the tachyon vacuum as 
expected~\cite{IU}. In the presence of both the rolling tachyon and 
a uniform electric field, there exists a critical value of the uniform 
electric field and perturbative waves on it propagate along electric 
flux lines, of which speed is given by magnitude of the background electric 
field~\cite{MS,GHY}. Though final remnants after decaying may be composed
of uniformly stretched fundamental strings (F1's) and tachyon matter, 
dynamical process including confinement is not understood yet.

An intriguing question worth asking at the present stage is to look for 
general homogeneous configuration of the effective field theory 
of the tachyon with a runaway potential and Born-Infeld type U(1) gauge
field, and to study properties of perturbative excitations on such
background. For the system of unstable D3-brane, we showed that 
the most general homogeneous solution is the time-dependent rolling tachyon 
with constant electric and magnetic fields. Under a reasonable shape of
the runaway tachyon potential, we find an exact form of the classical rolling 
tachyon solution, which may make further study more tractable
in the scheme of effective field theory. 
A family of tachyonic vacua constructed 
at infinite vacuum expectation value of the tachyon 
is characterized by magnitude of the constant background
electric and magnetic fields $(|{\bf E}_{0}|,|{\bf B}_{0}|)$ and the angle
$\theta$ between them. 
The configuration has nonvanishing momentum
along the direction perpendicular to ${\bf E}_0$ and ${\bf B}_0$.
By an appropriate Lorentz boost and a rotation we show that it is
a pressureless fluid of fundamental strings made by turning on 
electric and magnetic fields and
stretched along one direction in the rolling tachyon background,
which will be called string fluid with electric and magnetic fields.

There is however a subtle point. Since the momentum density is 
nonvanishing in the frame that fields are homogeneous, 
it is not the rest frame of the whole configuration. 
In the rest frame obtained by a suitable Lorentz boost, the tachyon
field is no longer homogeneous in the sense that $\partial_i T$ is not zero. 
In other words, the rest frame of the system 
and the frame where the tachyon field is homogeneous are different.
This fact makes the configuration rather nontrivial and in general the
electric field and the magnetic field are not completely (anti-)parallel 
though they form fluid of strings stretched along one direction.
The angle $\theta$
between them may then be interpreted as encoding the relative motion
between stretched strings along one direction and the rolling tachyon
flowing in the perpendicular direction to the strings. 

In order to reach the above interpretation we investigate the small 
fluctuations around the homogeneous configuration of the rolling
tachyon and the electromagnetic fields $(|{\bf E}_0|, |{\bf B}_0|, \theta)$.
After a Lorentz boost and a rotation, 
the fluctuations are shown to propagate along 
the direction of the electric field with the 
propagation velocity equal to the electric field. 
Roughly, the whole situation in this frame is then as follows. 
The tachyon has a nonzero momentum since it is Lorentz boosted; 
the electric field and the magnetic field still make a nonzero angle 
and have their own momentum. Then these two contributions cancel each other,
making the total momentum zero. This interpretation is also supported by
the fact that the only nonzero components of the energy-momentum tensor
are $T_{00}$ and the diagonal component along the direction of the electric
field.

This paper is organized as follows. In section 2, the most general 
homogeneous solution of rolling tachyon with constant electric and magnetic
field is obtained.
In section 3, propagation of small fluctuations on the obtained configuration
is analyzed. 
Then via a suitable Lorentz  boost and a rotation, we identify the
background configuration as a fluid of strings with electric and magnetic 
fields stretched along the direction of electric field and rolling tachyon.
In section
4, we conclude with discussion. Finally in appendix, detailed description 
of perturbative electromagnetic waves is given in the pure tachyon background.

\setcounter{equation}{0}
\section{Homogeneous Configuration}

In this section we introduce the system of tachyon coupled to an Abelian 
gauge field and find the most general homogeneous solution which
turns out to be constant electric and magnetic fields 
together with rolling tachyon configuration.

When we turn off antisymmetric tensor field of second rank in bosonic sector
of effective action, the unstable D3-brane system is described by
the following Born-Infeld type action~\cite{Gar,KMM}
\begin{equation}\label{fa}
S= -T_3 \int d^4x\; V(T) \sqrt{-\det (\eta_{\mu\nu} +
\partial_\mu T\partial_\nu T + F_{\mu\nu})}\, ,
\end{equation}
where $T$ is tachyon and $F_{\mu\nu}$ is field strength tensor of Abelian
gauge field $A_{\mu}$ on the D3-brane.

Since tachyon potential $V(T)$ measures varying tension, it should satisfy
two boundary values such that $V(T=0)=1$ and $V(T=\infty)=0$.
Specific computation based on (boundary) string field theory~\cite{GS,KMM} 
gives $V(T)\sim e^{-T^{2}}$ and Ref.~\cite{Sen} suggests
$V(T)\sim e^{-T}$ for large $T$.
Here we adopt a shape of the potential for the sake of convenient analysis 
\begin{equation}\label{V3}
V(T)=\frac{1}{{\rm cosh}\left(\frac{T}{T_{0}}\right)},
\end{equation}
where $T_{0}$ is determined by string theory of our interest.

To proceed, we introduce a few notations. We first define
\begin{eqnarray}
X_{\mu\nu}&\equiv & \eta_{\mu\nu} + \partial_\mu T\partial_\nu T + F_{\mu\nu},
\label{Xmn}\\
X&\equiv & \det (X_{\mu\nu}).
\label{X}
\end{eqnarray}
In $X_{\mu\nu}$, we separate barred metric $\bar{\eta}_{\mu\nu}$
and barred field strength tensor $\bar{F}_{\mu\nu}$
\begin{eqnarray}
\bar{\eta}_{\mu\nu}&=&\eta_{\mu\nu}+\partial_{\mu}T\partial_{\nu}T,
\label{emn}\\
\bar{F}_{\mu\nu} &=& F_{\mu\nu}.
\label{fmn}
\end{eqnarray}
Then we have determinant of barred metric
$\bar{\eta}$ and inverse metric
$\bar{\eta}^{\mu\nu}$
\begin{eqnarray}\label{eemn}
\bar{\eta} = -(1 + \partial_\mu T \partial^\mu T),~~~
\bar{\eta}^{\mu\nu} = \eta^{\mu\nu} -
\frac{\partial^\mu T \partial^\nu T}{1
+ \partial_\rho T \partial^\rho T},
\end{eqnarray}
and contravariant barred field strength tensor $\bar{F}^{\mu\nu}$ and
its dual field strength $\bar{F}^{\ast}_{\mu\nu}$
\begin{eqnarray}\label{ffmn}
\bar{F}^{\mu\nu} = \bar{\eta}^{\mu\alpha}
\bar{\eta}^{\nu\beta}F_{\alpha\beta},~~~
\bar{F}^{\ast}_{\mu\nu}= \frac{\bar{\epsilon}_{\mu\nu\alpha\beta}}
{2}
\bar{F}^{\alpha\beta}= \frac{\bar{\epsilon}_{\mu\nu\alpha\beta}}{2}
\bar{\eta}^{\alpha\gamma}\bar{\eta}^{\beta\delta}F_{\gamma\delta},
\end{eqnarray}
where $ \bar{\epsilon}_{\mu\nu\alpha\beta} = \sqrt{-\bar{\eta}}\;
\epsilon_{\mu\nu\alpha\beta}$ with $\epsilon_{0123} = 1$.

In terms of barred quantities 
Eq.~(\ref{X}) is computed as
\begin{eqnarray}\label{XX}
X = \bar{\eta} \left[ 1 + \frac12 \bar{F}_{\mu\nu}
\bar{F}^{\mu\nu} - \frac1{16} \left(\bar{F}^*_{\mu\nu}
\bar{F}^{\mu\nu}\right)^2 \right].
\end{eqnarray}
Then equations of motion for the tachyon $T$ and the gauge field
$A_{\mu}$ are
\begin{eqnarray}
\partial_\mu\left( \frac{V}{\sqrt{-X}}
\;C^{\mu\nu}_{\rm S}\; \partial_\nu T\right)
+\sqrt{-X}\; \frac{d V}{d T}
= 0,
\label{te} \\
\partial_\mu\left( \frac{V}{\sqrt{-X}}
\;C^{\mu\nu}_{\rm A}\right) = 0.
\label{ge}
\end{eqnarray}
Here $C^{\mu\nu}_{\rm S}$ and $C^{\mu\nu}_{\rm A}$ are symmetric
and asymmetric part, respectively, of the cofactor, 
\begin{equation}\label{cmn}
C^{\mu\nu} = \bar{\eta}\left(
\bar{\eta}^{\mu\nu} + \bar{F}^{\mu\nu} + \bar{\eta}^{\mu\alpha}
\bar{\eta}^{\beta\gamma}\bar{\eta}^{\delta\nu}
\bar{F}^*_{\alpha\beta}\bar{F}^*_{\gamma\delta}
+\bar{\eta}^{\mu\alpha}\bar{\eta}^{\beta\gamma}
\bar{F}^*_{\alpha\beta}\bar{F}^*_{\gamma\delta}\bar{F}^{\delta\nu}
\right),
\end{equation}
namely,
\begin{eqnarray}
C^{\mu\nu}_{\rm S} &=& \bar{\eta} (
\bar{\eta}^{\mu\nu}
+ \bar{\eta}^{\mu\alpha}
\bar{\eta}^{\beta\gamma}\bar{\eta}^{\delta\nu}
\bar{F}^*_{\alpha\beta}\bar{F}^*_{\gamma\delta}),\\
C^{\mu\nu}_{\rm A} &=& \bar{\eta}(\bar{F}^{\mu\nu}
+\bar{\eta}^{\mu\alpha}\bar{\eta}^{\beta\gamma}
\bar{F}^*_{\alpha\beta}\bar{F}^*_{\gamma\delta}\bar{F}^{\delta\nu}).
\end{eqnarray}

Energy-momentum tensor $T_{\mu\nu}$ in a symmetric form is given by
\begin{eqnarray}\label{tmne}
T_{\mu\nu} = - \frac{T_3 V(T)}{\sqrt{- X}}\;
\left[ -\eta_{\mu\nu} X
+ \frac{1}{2}\left( C_{\mu\rho}(\partial_\nu T \partial^\rho T
+ F_\nu^{~\rho}) +  C_{\nu\rho}(\partial_\mu T \partial^\rho T
+ F_\mu^{~\rho})\right)\right],
\end{eqnarray}
where $C_{\mu\nu}\equiv \eta_{\mu\alpha}\eta_{\nu\beta}
C^{\alpha\beta}$.
Diagonal components can be identified as density $\rho$
and averaged pressure $p$,
\begin{eqnarray}
\rho&=&  - \frac{T_3 V(T)}{\sqrt{- X}}\;
\left[ X +  C_{0\mu} (\dot{T}\partial^\mu T
+ F_0^{~\mu} )\right],  \label{rd}\\
p &=&  - \frac{T_3 V(T)}{3\sqrt{- X}}\;
\left[-3 X + C_{0\mu}(\dot{T}\partial^\mu T + F_0^{~\mu})
 + C^{\mu\nu}(\partial_\mu T\partial_\nu T + F_{\mu\nu})
\right]. \label{rnp}
\end{eqnarray}
{}From the equation of state $p=w \rho$, we read
\begin{eqnarray}
w &=& \frac13 - \frac{4 X - C^{\mu\nu}
(\partial_\mu T\partial_\nu T + F_{\mu\nu})}
{ 3\left[ X + C_{0\mu}(\dot{T}\partial^\mu T + F_0^{~\mu})
\right]}.\label{rw}
\end{eqnarray}
Thus trace of the energy-momentum tensor (\ref{tmne}) is given by
\begin{equation}\label{trt}
T^\mu_{~\mu} = -\frac {T_3 V(T)}{\sqrt{-X }} \left[ (4 + 3
\partial_\rho T \partial^\rho T) +\left((1+\frac12 \partial_\rho T
\partial^\rho T)\delta^\mu_{~\sigma} - \partial^\mu T \partial_\sigma T
\right) F_{\mu\nu} F^{\sigma\nu} \right].
\end{equation}

Let us consider homogeneous configurations of $T(t)$,
${\bf E}(t)$, and ${\bf B}(t)$ defined by $E^{i}=F_{i0}$ and 
$B^{i}=\epsilon_{0ijk}F_{jk}/2$. Then Faraday's law 
\begin{equation} \label{faraday}
\nabla\times{\bf E} + \frac{\partial{\bf B}}{\partial t} = 0
\end{equation}
allows only constant magnetic field ${\bf B}={\bf B}_{0}$. 
The equations of motion (\ref{te}) and (\ref{ge}) become
\begin{eqnarray}
&& \partial_0 \left( \frac{V}{\sqrt{-X}}\dot{T} 
\right) + \frac{\sqrt{-X}}{1+{\bf B}_{0}^{2}} \frac{dV}{dT} = 0, 
\label{homeq} \\
&& \left(\delta_{ij}+B_{0}^{i}B_{0}^{j}\right)\partial_0 
\left( \frac{V}{\sqrt{-X}} E^{j}\right) = 0,
\label{emza}
\end{eqnarray}
where  
\begin{eqnarray}\label{XX1}
X 
= -(1-\dot T^2) ( 1+ {\bf B}^2) + {\bf E}^2 
+({\bf E}\cdot {\bf B})^2.
\end{eqnarray}
The energy density $\rho$ (\ref{rd}) reduces to
\begin{equation}\label{enden}
\rho = T_{3}\frac{V}{\sqrt{-X}} ( 1 + {\bf B}_0^2),
\end{equation}
and is a constant, which can be seen from the equations 
(\ref{homeq})--(\ref{emza}) as it should be.
Eq.~(\ref{emza}) then becomes
\begin{equation}\label{emza1}
(\delta^{ij} + B_0^i B_0^j)\dot{E}^{j}=0.
\end{equation}
Since the matrix $\delta^{ij} + B_0^i B_0^j$ is positive definite,
Eq.~(\ref{emza1}) allows only constant electric field $ E^i = E_{0}^{i}.$
Therefore, the most general homogeneous solution of 
Eqs.~(\ref{homeq})--(\ref{emza}) is ${\bf E}={\bf E}_{0} = \mbox{constant}$, 
${\bf B}={\bf B}_{0} = \mbox{constant}$, and $V/\sqrt{-X}= \mbox{constant}$. 
When ${\bf E}_{0}$ and ${\bf B}_{0}$ are not parallel or anti-parallel
to each other, this configuration has nonvanishing linear momentum
density 
\begin{equation} \label{mom}
{\cal P}_i = T^{0i} = T_3 \frac{V}{\sqrt{-X}} \epsilon_{ijk} E_0^j B_0^k,
\end{equation}
so does angular momentum density.

Since $\rho$ is a constant, $X$ must vanish at the minimum of the tachyon 
potential $V(T\rightarrow\infty)=0$, and it determines
$\dot T_\infty=\dot T(T\rightarrow\infty)$
\begin{equation}\label{vacf1}
\dot T_\infty^{2}= 1-\frac{{\bf E}_{0}^{2} + 
({\bf E}_0\cdot {\bf B}_0)^2}{1+{\bf B}_{0}^{2}}.
\end{equation}
Thus, the solution is specified by three parameters,
namely, the magnitudes of ${\bf E}_0$ and ${\bf B}_0$,
and the angle $\theta$ between them.
When ${\bf B}_{0}=0$, Eq.~(\ref{vacf1}) coincides trivially 
with the result of pure electric case in Ref.~\cite{MS,GHY}. 
If ${\bf E}_{0}=0$, the presence of
${\bf B}_{0} \ne 0$ plays no role in the equation of motion
(\ref{homeq}) except for a constant overall scale.
When ${\bf E}_{0}\ne 0$ and ${\bf B}_{0}\ne 0$, 
we consider the set of allowed values of $(|{\bf E}_{0}|,|{\bf B}_{0}|,\theta)$
for given initial values $T_{i}$ and $\dot{T}_{i}$ 
of $T$ and $\dot{T}$, satisfying $-X>0$. Then, the energy density (\ref{enden})
has maximum
for both parallel $(\theta=0)$ and anti-parallel $(\theta=\pi)$ cases,
and minimum for orthogonal case $(\theta=\pi/2)$.
For a given angle $\theta$,
the energy density $\rho$ is monotonically-increasing function of both 
$|{\bf E}_{0}|$ and $|{\bf B}_{0}|$ as expected. 
When ${\bf E}_{0}$ is parallel or anti-parallel to ${\bf B}_{0}$,
$|{\bf E}_{0}|$ has upper bound $\sqrt{1-\dot{T}_{i}^{2}}$ irrespective 
of $|{\bf B}_{0}|$.
When ${\bf E}_{0}$ is orthogonal to ${\bf B}_{0}$, upper limit of 
$|{\bf E}_{0}|$ is increased as $|{\bf B}_{0}|$ increases so that
a configuration of infinite $|{\bf E}_{0}|$ and $|{\bf B}_{0}|$ satisfying
${\bf E}_{0}^{2}-{\bf B}_{0}^{2}=-F^{0}_{\mu\nu}F_{0}^{\mu\nu}/2\le 1$ 
is available.

One may wonder whether the solutions ($|{\bf E}_{0}|,|{\bf B}_{0}|,\theta$)
are not all independent but related by some Lorentz transformations which
could change the angle $\theta$, for example. 
However this is not the case since, under a Lorentz boost, the tachyon 
field $T$ is no longer homogeneous in space. Therefore in general
they describe physically distinct configurations. To be  more explicit,
let us work in the coordinate where the electric field is directed in the 
$x$-direction and the magnetic field lies on the $xy$-plane,
\begin{equation} \label{e1b12}
{\bf E}_0 = E_0 \hat{\bf x}, \qquad 
{\bf B}_0 = B_1 \hat{\bf x} + B_2 \hat{\bf y},
\end{equation} 
where $B_1=|{\bf B}_0|\cos\theta$ and $B_2=|{\bf B}_0|\sin\theta$.
This configuration has the momentum density ${\cal P}_3$ in the 
$z$-direction as seen in Eq.~(\ref{mom}). Other nonvanishing components of 
the energy-momentum tensor (\ref{tmne}) are
\begin{eqnarray} \label{emtensor}
&&T_{11} = - T_3 \frac{V}{\sqrt{-X}}(1-\dot T^2)(1+B_1^2), \nonumber \\
&&T_{22} = - T_3 \frac{V}{\sqrt{-X}}\left[
         (1-\dot T^2)(1+B_2^2) - E_0^2 \right], \nonumber \\
&&T_{33} = - T_3 \frac{V}{\sqrt{-X}} ( 1-\dot T^2 - E_0^2 ), \nonumber\\
&&T_{12} = - T_3 \frac{V}{\sqrt{-X}}(1-\dot T^2)B_1 B_2.
\end{eqnarray}
We will discuss more about this homogeneous configuration in section 3 when 
small fluctuations around it are considered.

Let us turn to rolling of the tachyon condensate
in the presence of uniform electric and magnetic fields.
For the tachyon potential (\ref{V3}),
from constancy of the energy density, we have a solution
\begin{eqnarray}\label{ts}
T(t) = T_0\sinh^{-1}\left(
    a_+e^{(\dot T_\infty/T_0)\,t}-a_-e^{-(\dot T_\infty/T_0)\,t}\right),
\end{eqnarray}
where $a_\pm=\frac12\left[(\dot T_i/\dot T_\infty)\cosh(T_i/T_0)
\pm\sinh(T_i/T_0)\right]$.
Compared to the case of no electromagnetic field
for which $\dot T_\infty=1$,
the solution (\ref{ts}) has rescalings of the energy density
$\rho \rightarrow
\rho\left[(1+{\bf B}_0^2)(1-\dot T_i^2)/(\dot T_\infty^2-\dot T_i^2)\right]^{1/2}$
and $T_0 \rightarrow T_0/\dot T_\infty$
in the exponents of Eq.~(\ref{ts}).
As the value of $[{\bf E}_0^2+({\bf E}_0\cdot{\bf B}_0)^2]/(1+{\bf B}_0^2)$
approaches unity, $\dot T_\infty$ almost vanishes from Eq.~(\ref{vacf1})
and not only the energy density (\ref{enden}) becomes larger, but also
the time scale of remaining on top of the tachyon potential becomes longer.
This may be a nice feature for various aspects in cosmology, e.g.,
the rolling tachyon as a source of sufficient
inflation~\cite{Gib,KL,tacos1,KKK,MW}.
In trials to use pure tachyon rolling as a source of inflation,
sufficient inflation could not be obtained with the potential
derived from string theory and {\it ad hoc} potentials had to be introduced.
With the tachyon in the gauge field background, sufficient inflation
may be obtained near the top of the tachyon potential.
However, a difficulty in applying unstable D-branes with electric
and magnetic flux to cosmology is that the constant electric and magnetic
field background is homogeneous, but not isotropic.
Therefore, the electric and magnetic fields must disappear at the end
of inflation.

\setcounter{equation}{0}
\section{Small Fluctuations and Electromagnetic String Fluid}

In this section, we investigate the propagation of small fluctuations
in the homogeneous background field discussed in the previous section. 
Through the analysis we will argue that the homogeneous configuration
is identified as a fluid consisting of strings carrying electric and
magnetic fields stretched along one
direction and the rolling tachyon flowing in a perpendicular direction
to the strings. The relative velocity between strings
and the rolling tachyon is essentially encoded in the angle $\theta$ 
between the electric field and the magnetic field.

In the pure rolling tachyon background,
propagation of small fluctuations is completely suppressed~\cite{IU},
meaning the absence of perturbative degrees of freedom.
In Ref.~\cite{MS,GHY}, it was shown that with the electric field turned on
1+1 dimensional propagation of small fluctuations is allowed along
the electric flux lines.
Here we first extend this analysis including the magnetic field in 
unstable D3-branes.

In the rolling tachyon limit $\dot T\rightarrow\dot T_\infty$,
with the constant electric and magnetic field background,
the equations of motion of the gauge fields and the tachyon are combined 
into\footnote{We neglected the potential part since we are mainly 
interested in the causal behavior \cite{GHY}.}
\begin{equation}
\label{feq}
(\eta+F)^{-1\ (AB)} \partial_AF_{BC}=0 ,
\end{equation}
where indices $A,B,...=0,...,3,T$
with $F_{\mu T}=-F_{T\mu}=\partial_\mu T$
and $(AB)$ means the symmetric part of $(\eta+F)^{-1}$. 
The fields are divided into the background $F^0_{AB}$
and the small fluctuations $f_{AB}$
\begin{equation}
F_{AB}=F^0_{AB}+f_{AB}.
\end{equation}
We choose the frame where the background fields are homogeneous and
work in the background  (\ref{e1b12}) considered in section 2.
Then the background $F^0$ takes the form
\begin{equation}
F^0 = \left(\begin{array}{ccccc}
 0 & -E_0 & 0 & 0 & \dot T \\
 E_0 & 0 & 0 & -B_2 & 0 \\
 0 & 0 & 0 & B_1 & 0 \\
 0 & B_2 & -B_1 & 0 & 0 \\
-\dot T & 0 & 0 & 0 & 0
\end{array}\right).
\end{equation}

In the rolling tachyon limit,
$\dot T=\dot T_\infty=\sqrt{1-E_0^2(1+B_1^2)/(1+{\bf B}_0^2)}$.
The fluctuations $f_{AB}$ can be written in terms of the gauge potential $a_\mu$
and the tachyon fluctuation $\hat{t}$ as
$f_{\mu\nu}=\partial_\mu a_\nu-\partial_\nu a_\mu$ and
$f_{\mu T}=\partial_\mu \hat{t}$.
We will work in Weyl gauge $a_0=0$ and define
\begin{eqnarray}
\partial'_0 &=& \frac{(1+{\bf B}_0^2)\partial_0+E_0B_2\partial_3}
    {\sqrt{(1+{\bf B}_0^2)^2-E_0^2B_2^2}}, \nonumber\\ 
\partial'_1 &=& \frac{(1+B_1^2)\partial_1+B_1B_2\partial_2}
    {\sqrt{(1+B_1^2)^2+B_1^2B_2^2}}, \nonumber\\
{v'}^2 &=& \frac{E_0^2\left[(1+B_1^2)^2+B_1^2B_2^2\right]}
    {(1+{\bf B}_0^2)^2-E_0^2B_2^2} \le 1,
\end{eqnarray}
where the inequality in the last line comes from Eq.~(\ref{vacf1}).
Then the equations of motion of the fluctuations derived from 
Eq.~(\ref{feq}) are
\begin{eqnarray}
&& E_0\partial_0 G = 0, \label{gf}\\
&&\partial'^2_0 a_i - {v'}^2\partial'^2_1 a_i
     - \frac{E_0}{(1+{\bf B}_0^2)^2 - E_0^2 B_2^2}\partial_i G = 0, \\
&& \partial'^2_0 \hat{t} - {v'}^2\partial'^2_1 \hat{t} = 0,
\end{eqnarray}
where 
\begin{eqnarray}
G &=& \sqrt{(1+B_1^2)^2+B_1^2B_2^2}\; \partial'_1 \left[
\sqrt{(1+{\bf B}_0^2)^2-E_0^2(1+{\bf B}_0^2)(1+B_1^2)} \hat{t}
\right.\nonumber\\ && \left.\vphantom{\sqrt{(1+{\bf B}_0^2)^2}}
- E_0\left((1+B_1^2)a_1+B_1B_2a_2\right) \right]
+ \sqrt{(1+{\bf B}_0^2)^2-E_0^2B_2^2}\; B_2\partial'_0 a_3.
\end{eqnarray}
The first equation (\ref{gf}) implies $G=G({\bf x})$.
Choice of the $a_0=0$ gauge leaves us the space-dependent gauge transformation
$\Lambda({\bf x})$ and under this gauge transformation $G({\bf x})$ 
transforms to
$G({\bf x})-E_0\left[\left((1+B_1^2)\partial_1+B_1B_2\partial_2\right)^2
-B_2^2\partial_3^2\right]\Lambda({\bf x})$.
Thus $G({\bf x})$ is not a propagating mode and we can set $G({\bf x})=0$
as a gauge choice.
Now there remain three propagating modes
which are mixtures of $a_i$ and $\hat{t}$,
and obey the wave equation of the form
\begin{equation}\label{wvq}
\partial'^2_0a-{v'}^2\partial'^2_1a=0.
\end{equation}
Therefore, the fluctuations propagate along only one direction,
$x$-direction in the primed frame,
and $v'$ is the propagating speed in that frame.
The Lorentz transformation connecting the homogeneous frame to
the primed frame is composed of a rotation in the $xy$-plane by 
\begin{equation}\label{varp}
\varphi=\tan^{-1}\left(\frac{B_1B_2}{1+B_1^2}\right)
=\tan^{-1}\left(
    \frac{{\bf B}_0^2\cos\theta\sin\theta}{1+{\bf B}_0^2\cos^2\theta}\right)
\end{equation}
and a boost along $z$-direction by $\beta=E_0B_2/(1+{\bf B}_0^2)$. 
Propagation velocity ${\bf v}$ in the homogeneous frame is given by 
\begin{equation}\label{hvel}
{\bf v} =
(\sqrt{1-\beta^2}\,v'\cos\varphi,\sqrt{1-\beta^2}\,v'\sin\varphi,\beta),
\end{equation}
and the speed is
\begin{equation}
v^2 = \beta^2 + (1-\beta^2)\,{v'}^2=
\frac{{\bf E}_0^2 + ({\bf E}_0 \cdot  {\bf B}_0)^2}{1+{\bf B}_0^2}
= 1-\dot T_\infty^2.
\end{equation}
Note that it is nothing but an expression in the vacuum condition
(\ref{vacf1}). 
For fixed $E_0$ and ${\bf B}_0$, the velocity $v$ has the maximum value $E_0$
when ${\bf E}_0$ and ${\bf B}_0$ are parallel or anti-parallel, and
the minimum value $E_0/\sqrt{1+{\bf B}_0^2}$ when they are orthogonal.
The propagation direction angle $\varphi$ in the $xy$-plane increases
as the angle $\theta$ between the electric and magnetic fields increases
from zero, reaching a maximum value
$\varphi_{\rm max}=\tan^{-1}({\bf B}_0^2/2\sqrt{1+{\bf B}_0^2})$
at $\theta=\cos^{-1}(1/\sqrt{2+{\bf B}_0^2})$,
then decreases back to zero at $\theta=\pi/2$.
For any value of $|{\bf B}_0|$,
we have $\varphi<\theta$ for $0<\theta\leq\pi/2$.
For $\pi/2<\theta<\pi$,
$\varphi$ takes a negative value $-\varphi(\pi-\theta)$ which keeps
$|\varphi|<\theta-\pi/2$.
The direction of propagation coincides with that of $|{\bf B}_0|$
in the limit $|{\bf B}_0|\cos\theta\rightarrow\infty$.

Since the wave equation of the fluctuations has the standard form (\ref{wvq})
in the primed frame, it would be illuminating to reconsider the configuration
in the primed frame. After the Lorentz boost and the rotation considered
above, background gauge fields are transformed to
\begin{eqnarray} \label{ebprime}
{\bf E}'_{0} &\equiv & E'_0 \hat{{\bf x}}' 
             = v' \hat{{\bf x}}' ,\nonumber\\
{\bf B}'_{0} &\equiv & B'_1 {\hat {\bf x}}' + B'_2 \hat{{\bf y}}' \nonumber\\
         & = & \frac{E_0}{v'} \left[ B_1 \hat{{\bf x}}'
        + \frac{\dot T_\infty^2 (1+{\bf B}_0^2)}%
               {(1+{\bf B}_0^2)^2 - E_0^2 B_2^2} B_2 \hat{{\bf y}}' \right] .
\end{eqnarray}
Also the energy-momentum tensor in the primed frame can be obtained from
Eq.~(\ref{emtensor}),
\begin{eqnarray} \label{emtensor2}
&&T'_{00} =  T_3 \frac{V}{\sqrt{-X}} \left[
   \frac{(1+{\bf B}_0^2)^2 - E_0^2 B_2^2}{1+{\bf B}_0^2}
  -\frac{E_0^2 B_2^2}{(1+{\bf B}_0^2)^2 - E_0^2 B_2^2}
   (\dot T_\infty^2 - \dot T^2) \right], \nonumber \\
&&T'_{11} = -v'^2 T'_{00} 
   - T_3 \frac{V}{\sqrt{-X}} 
   \frac{(1+B_1^2)^3 + B_1^2 B_2^2 (2B_1^2 + B_2^2 + 3)}%
        {(1+B_1^2)^2 + B_1^2 B_2^2} (\dot T_\infty^2 - \dot T^2), \nonumber \\
&&T'_{03} =  -T_3 \frac{V}{\sqrt{-X}} 
   \frac{E_0 B_2 (1+{\bf B}_0^2)}{(1+{\bf B}_0^2)^2 - E_0^2 B_2^2} 
   (\dot T_\infty^2 - \dot T^2), \nonumber \\
&&T'_{22} =  -T_3 \frac{V}{\sqrt{-X}} 
   \frac{(1+B_1^2) (1+{\bf B}_0^2)}{(1+B_1^2)^2 + B_1^2 B_2^2}
   (\dot T_\infty^2 - \dot T^2), \nonumber \\
&&T'_{33} =  -T_3 \frac{V}{\sqrt{-X}} 
   \frac{(1+{\bf B}_0^2)^2}{(1+{\bf B}_0^2)^2 - E_0^2 B_2^2}
   (\dot T_\infty^2 - \dot T^2), \nonumber \\
&&T'_{12} =  -T_3 \frac{V}{\sqrt{-X}} 
   \frac{B_1 B_2 (1+{\bf B}_0^2)}{(1+B_1^2)^2 + B_1^2 B_2^2}
   (\dot T_\infty^2 - \dot T^2). 
\end{eqnarray}
Then, in the limit $\dot T \rightarrow \dot T_\infty$, the only nonvanishing 
components are $T'_{00}$ and $T'_{11}$.

Interpretation of the configuration is now more clear in the
primed frame. Note that $T'_{22}$ and $T'_{33}$ vanish and hence there is 
no pressure in $y'$ and $z$ directions; the final configuration of the
rolling tachyon limit may then be 
considered as a pressureless fluid of one-dimensional objects stretched along 
$x'$-direction. Also, the momentum density 
$T'^{0i}$ vanishes in this frame, meaning that this frame corresponds
to the rest frame of the string fluid background which
consists of the homogeneous gauge fields and the rolling tachyon. 
Of course this is consistent with the fact that the equations for small 
fluctuations have the standard form (\ref{wvq}) in the primed frame. 
Moreover, the propagation velocity ${\bf v}'$ is nothing but
the same as the electric field ${\bf E}'_0$ in the rest frame of the fluid.

When there is no magnetic field  ${\bf B}_0 = 0$, such a one-dimensional
object is often identified as the fundamental string~\cite{DGHR,Yi,MS,GHY}. 
In the 
present case that ${\bf B}_0 \neq 0$, one may also interpret the object 
as a sort of string with electric and magnetic fields though more 
detailed string-theory analysis including confinement mechanism is needed
to draw any definite conclusion.  
Note, however, that ${\bf E}'_0$ and 
${\bf B}'_0$ are not in general (anti-)parallel in Eq.~(\ref{ebprime}). 
This can be understood roughly in the following way. In the 
Lorentz-boosted primed
frame, the tachyon field is no longer homogeneous in the sense that
$\partial_i T$ is not zero but it has linear momentum
in $z$-direction, $T = T((t'-\beta z)/\sqrt{1-\beta^2})$.
Then the contribution of the gauge field should cancel 
the momentum of the tachyon since total momentum density is zero.
This is possible if the electric and the magnetic field make a
nonzero angle, which is realized by boosting parallel electric
and magnetic field configurations. This interpretation is also 
consistent with the facts that $B'_2$ vanishes when $\dot T_\infty$ 
goes to zero in Eq.~(\ref{ebprime}).

Now the whole picture of the rolling tachyon in electromagnetic background
is as follows. Initially the unstable D3-brane
has constant electric and magnetic fields with an arbitrary angle. 
As brane decays the electric field and the magnetic field
form a pressureless fluid of strings.
Also, when they are not parallel, part of
the energy is used to boosts the strings relative to the homogeneous 
rolling tachyon field, rather than the entire energy is exhausted in
the formation of string-like object or lower dimensional D-branes.
Though the brane-decaying process would occur rather
violently instead of occurring through homogeneous tachyon rolling,
one might still think of the expression of the energy-momentum tensor 
in Eq.~(\ref{emtensor2}) as roughly representing how the formation and boost
of the string fluid is realized as $\dot T$ approaches $\dot T_\infty$. 

The final configuration is then 
a fluid consisting of string with electric and magnetic fields stretched 
along $x'$-directions and the rolling tachyon
moving in $z$-direction relative to the strings. The rest frame of the
whole configuration is the primed frame defined above in which small 
fluctuations propagate along the $x'$-direction with speed 
$v'=|{\bf E}'_0|$. Note however that the fluid is not a noninteracting sum of 
strings and the rolling tachyon; it is rather a nontrivial composite, since 
the gauge fluctuations and the tachyon fluctuation do not separate but are 
mixed together as seen above. Since this is due to the
effect of inhomogeneous configuration of the tachyon field in the rest
frame of the fluid, it would be interesting to study the consequence of
more nontrivial inhomogeneous rolling tachyon configurations~\cite{Senop}.  

\setcounter{equation}{0}
\section{Conclusion and Discussion}

We have seen that effective field theory of unstable D3-brane system, composed
of a tachyon field and an Abelian gauge field, supports time-dependent 
rolling tachyon configuration with constant electric and magnetic fields as
the most general homogeneous solution. Tachyonic vacua are labeled by 
magnitudes of electric and magnetic fields and the angle between them. 
Through the analysis of small fluctuations in this background we find
that the final configuration in the rolling tachyon limit is interpreted as 
a fluid consisting of strings with electric and magnetic fields stretched 
along one direction and the rolling tachyon moving
in a perpendicular direction to the strings. The fluctuations propagate
along the direction of strings with velocity equal to the electric field.
 
Though we considered the case of unstable D3-brane, it would be interesting 
to understand D$p$-branes of arbitrary dimensions. 
For higher dimensional D$p$-branes, further study is needed because
determinant in the Born-Infeld type action 
contains higher-order terms and magnetic 
components of the field strength tensor become more complicated 
such as $B_{i_{1}i_{2}\cdots i_{p-2}}=
\epsilon_{0i_{1}i_{2}\cdots i_{p}}F_{i_{p-1}i_{p}}/2$. 
However, once the objects 
are generated, then they are likely to be higher dimensional analogue
of the strings with electric and magnetic fields.
When $p=2$, magnetic field has single component $B=\epsilon_{012}F_{12}$ 
and the determinant in the Born-Infeld type action is quadratic, and hence
only the orthogonal case is realized.
This kind of analysis should also be addressed in the scheme of string
field theory~\cite{Sen,MS,Senop,ST}.

It would be interesting if the following points are also understood.
Though our analysis has been accomplished by adopting the specific action from
Ref.~\cite{Gar,Sen,GS}, vacuum structure and propagation of electromagnetic 
field probably share universal 
behavior even in general BSFT type action~\cite{KMM} due to the simple 
limiting property of kinetic term of the BSFT actions as argued in 
Ref.~\cite{IU,GHY}. 
In section 3 and appendix, we utilized quadratic expansion for fluctuations
of the tachyon and the gauge field to identify mainly stringy objects
and their propagating velocity, however
description of mutual interaction between fluctuating degrees
requires at least quartic expansion or
some nonperturbative approaches.

Addition of other bosonic degrees are also worth tackling. Inclusion of 
gravity may let us ask what the gravitating version of the stretched 
strings in the rolling tachyon background is.
If we take into account the rolling tachyon as a viable source of inflationary 
universe, then photons of Born-Infeld type action on the unstable branes
can also be testified as a component of materializing radiation dominated era.
In such sense existence of constant electromagnetic background field
seems beneficial for boosting generation of the photons even in late time,
but direct application is difficult due to the lack of isotropy. 
Taking into account gravitational radiation is also intriguing to 
understand the fate of tachyon matter~\cite{CLL}.
When we can deal with constant background antisymmetric tensor 
field $B_{\mu\nu}$ which is the same as constant background electromagnetic 
field $({\bf E}_{0},{\bf B}_{0})$ in the action on the brane, distinction
between them can easily be tested by pattern of propagating small
fluctuations in the given background.

\vspace{9mm}

\noindent {\large {\bf Note Added}}

\noindent After completion of this work, Ref.~\cite{RS} appeared on the 
archive, which considered the case that electric and magnetic fields are 
orthogonal in the context of both string and effective theories. 

\setcounter{equation}{0}
\section*{Acknowledgments}
We would like to thank Piljin Yi, Hyeonjoon Shin, 
and Jeong-Hyuck Park for helpful discussions.
H.B.~Kim would like to acknowledge the hospitality
of Korea Institute for Advanced Study where a part of this work has been done.
Y.~Kim would like to thank E-Ken of Nagoya University for its hospitality,
where a part of this manuscript was written. 
This work was supported by Korea Research Foundation Grant
KRF-2002-070-C00025(C.K.) and the Swiss Science Foundation,
grant 21-58947.99(H.B.K.), and is
the result of research activities (Astrophysical Research
Center for the Structure and Evolution of the Cosmos (ARCSEC) and
the Basic Research Program, R01-2000-000-00021-0)
supported by Korea Science $\&$ Engineering Foundation(Y.K. and O.K.).

\appendix

\section{Electromagnetism in the Pure Tachyon Background}

In this appendix we will discuss in detail 
the fluctuations around rolling tachyon
background (\ref{ts}) without electric or magnetic field, ${\bf E}_{0}=
{\bf B}_{0}=0$.

The fluctuation $\delta T(t,{\bf X})$ of tachyon satisfies the
linearized equation
\begin{equation}\label{dte}
\partial_0\left( \frac{\delta \dot T}{1-\dot T^2} \right) 
        - \nabla^2 \delta T 
+\frac{d^2\ln V}{dT^2} \delta T = 0,
\end{equation}
and it decouples from the electromagnetic 
fluctuations. Near the top of the potential it reduces to the Klein-Gordon 
equation with negative mass and diverges exponentially as it should.
At late time, $\dot T \rightarrow 1$, the first term dominates and the
fluctuation stops propagating~\cite{Sen}. 
As we will see shortly, the fluctuation of
the gauge field also shows the same behavior. Actually this is a consequence
of Carrollian limit of vanishing light speed at late time of
rolling tachyon without background electromagnetic field~\cite{IU,MS,GHY}.

Now we consider the propagation of the gauge field $A_{\mu}$ in the 
rolling tachyon background (\ref{ts}) by linearizing the
equation of motion (\ref{ge}) in this background. Written in terms
of electric and magnetic fields, it becomes a modified Amp\'{e}re's law
\begin{equation}\label{ameq}
\nabla\times {\bf B}=\frac{\epsilon_{0}^{2}}{V(t)^{2}}
\frac{\partial {\bf E}}{\partial t},
\end{equation}
where we have used the fact that
$V(T)/\sqrt{1-\dot{T}^{2}}\equiv\epsilon_{0}$ is a constant.
Then one may identify the tachyon potential $V(t)/\epsilon_0$ 
as the varying speed of electromagnetic waves. In particular, the late time 
behavior $V(t)\rightarrow 0$ corresponds to the Carrollian limit. 
This seems to suggest a dictionary `velocity of electromagnetic waves 
and tachyon potential' like `time and tachyon' in Ref.~\cite{Sen2}.
However, it should be confirmed by careful analysis.

Combined with the Faraday's law (\ref{faraday}), the wave equations of the 
electric field ${\bf E}$ and the magnetic field ${\bf B}$ can be written as
\begin{eqnarray}
\frac{\epsilon_{0}^2}{V^2}\frac{\partial^2{\bf B}}{\partial t^2}
-\nabla^2{\bf B} &=& 0,\label{beq1}\\
\frac{\partial}{\partial t} \left(\frac{\epsilon_{0}^2}{V^2}
\frac{\partial {\bf E}}{\partial t} \right)
-\nabla^2{\bf E} &=& 0. \label{eeq2r}
\end{eqnarray}
Note that these equations are not symmetrical since the Amp\'{e}re's law
(\ref{ameq}) is modified but Faraday's law (\ref{faraday}) is left unchanged.
More specifically, Eq.~(\ref{eeq2r}) has a kind of damping term which is 
the first order in time derivative and the propagation of electric field 
shows rather different behavior from that of magnetic field as demonstrated 
below.

To investigate in detail the propagation of electromagnetic wave in the
rolling tachyon background, let us try to find a plane-wave solution of 
Eqs.~(\ref{beq1}) and (\ref{eeq2r}).
Without loss of generality we consider the plane-wave propagating along 
the $x$-axis,
${\bf k}=k\hat{\bf x}$, and make an ansatz ${\bf B}=B_z(t,x)\hat{\bf z}$ with
\begin{equation} \label{Ba}
B_z(t,x) = B(t)e^{ik[x-W_B(t)]},
\end{equation}
where $B(t)$ and $W_B(t)$ are real.
We assume that $V(0)\approx1$ and $\dot W_B(0)\approx1$ so that
the ansatz (\ref{Ba}) gives the normal plane-wave initially.
Since $\bf E$ and $\bf B$ are related through the Amp\'ere's law 
(\ref{ameq}) and Faraday's law (\ref{faraday}),
${\bf E} = E_y(t,x) \hat{\bf y}$ is then given by
\begin{equation}
E_y(t,x) = E(t)e^{ik[x-W_E(t)]},
\end{equation}
with
\begin{eqnarray}
E &=& \sqrt{\dot W_B^2B^2+k^{-2}\dot B^2}, \label{e-b1}\\
W_E &=&  W_B - k^{-1}\tan^{-1}\left(\frac{\dot B}{k \dot W_BB}\right).
\label{e-b2}
\end{eqnarray}
With this ansatz, Eq.~(\ref{beq1}) reduces to
\begin{eqnarray}
\ddot B-k^2\left(\dot W_B^2-\epsilon_{0}^{-2}V^2\right)B &=& 0, 
\label{fir}\\
2\dot B\dot W_B + B\ddot W_B &=& 0. \label{sec}
\end{eqnarray}
Eq.~(\ref{sec}) implies 
\begin{equation}
\dot W_B = \frac{1}{\epsilon_{0}}\left(\frac{B_0}{B}\right)^2, \label{Bspeed}
\end{equation}
where $B_0$ is a constant. From Eq.~(\ref{e-b2}) we also find
\begin{equation} \label{Espeed}
\dot W_E = \frac{V^2}{\epsilon_{0}^3}\left(\frac{B_0}{E}\right)^2.
\end{equation}
Inserting Eq.~(\ref{Bspeed}) into Eq.~(\ref{fir}), we obtain a
second-order nonlinear equation for amplitude of the magnetic field
\begin{equation}
\ddot B - \frac{k^2}{\epsilon_{0}^2}\left[
   \left(\frac{B_0}{B}\right)^4 - V^2 \right] B = 0. 
\label{Beq}
\end{equation}
On the top of the potential $V=1$, we get a normal plane-wave solution
$\dot W_B=\dot W_E=1/\epsilon_{0}$ with amplitudes of $B = B_0$ and 
$E=B_0/\epsilon_{0}$ as expected from the original wave equations 
(\ref{beq1})--(\ref{eeq2r}). 

\begin{figure}
\centerline{\epsfig{figure=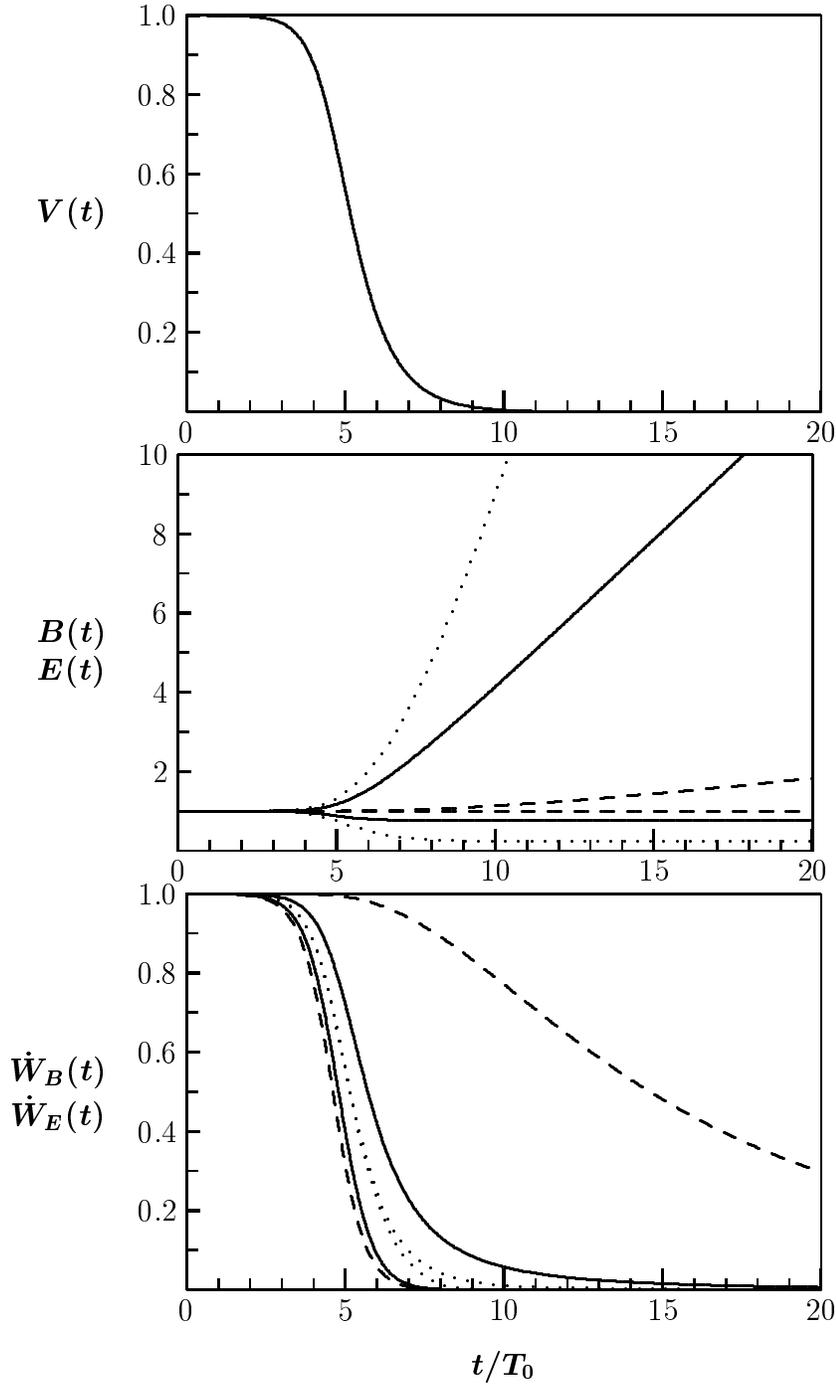}}
\caption{
(Top) tachyon potential as a function of time, Eq.~(\ref{vtt})
for $a_+=0.1$ and $a_-=0$ ($\epsilon_0=1$).
(Middle) amplitudes and (bottom) propagation speeds of
magnetic field (upper three lines) and electric field (lower three lines)
for $kT_0=0.1$ (dashed line), $kT_0=1$ (solid line),
and $kT_0=10$ (dotted line).
}
\label{fig1}
\end{figure}

To find more detailed behavior of the solution, we need to know the 
explicit form of $V(T(t))$. With the tachyon potential (\ref{V3}) and
the tachyon background (\ref{ts}), it becomes
\begin{equation}\label{vtt}
V(T(t)) =\frac{1}{\sqrt{1+\left(a_+e^{t/T_0}-a_-e^{-t/T_0}\right)^2}}.
\end{equation}
In this case we solved Eq.~(\ref{Beq}) numerically and the result is shown
in  Fig.~\ref{fig1}. Note that the electric field goes to a constant value,
while the magnetic field linearly diverges. Also both $\dot W_E$ and 
$\dot W_B$ vanish after some time, which means that electromagnetic wave
freezes and does not propagate after the tachyon rolling.
In addition, one can see from the last figure of Fig.~\ref{fig1}
that waves with short 
wavelengths are frozen earlier than those with long wavelengths.
Finally $\dot W_E$ tends to
vanish faster than $\dot W_B$, i.e., the electric field is frozen more
quickly than the magnetic field. This is because $\dot W_E$ in (\ref{Espeed})
contains extra $V^2$ compared with $\dot W_B$ in Eq.~(\ref{Bspeed}).

For the onset behavior, one can also obtain analytical expressions.
Once the tachyon rolls from the top of the potential, the fluctuations of
the gauge fields start deviating from the normal plane-wave solution.
With the initial condition $V(t=0)=1$, that is $T_i=0$, $V(T(t))$ reduces to
\begin{equation}
V(T(t)) = \frac{1}{\sqrt{1+\dot T_i\sinh^2(t/T_0)}}.
\end{equation}
Assuming that $\dot T_i$ is small,
which means the tachyon rolls slowly initially,
we can solve Eqs.~(\ref{Beq}) and (\ref{Bspeed})
to the first-order in $\dot T_i^2$:
\begin{eqnarray}
B(t) &=& B_0 \left[1 + \frac{\dot T_i^2\,k^2 T_0^2}{4(1+k^2 T_0^2)}
    \sinh^2\left(\frac{t}{T_0}\right)\right], \nonumber \\
W_B(t) &=& t - \frac{\dot T_i^2}{4}\left[
    t+\frac{k^2 T_0^3}{2(1+k^2T_0^2)}\sinh\left(\frac{2t}{T_0}\right)\right], 
\end{eqnarray}
where we have rescaled $B_0$ so that $B(0)=B_0$.
This solution is valid as long as $\dot T_i^2e^{t/T_0}\ll1$.
Thus the amplitude initially grows quadratically as the tachyon starts rolling.
On the other hand, electric field can be calculated
from Eqs.~(\ref{e-b1}) and (\ref{Espeed}),
\begin{eqnarray}
E(t) &=& B_0 \left[1 - \frac{\dot T_i^2}{4(1+k^2T_0^2)}
    \left(1+2k^2T_0^2+k^2T_0^2\sinh^2\left(\frac{t}{T_0}\right)\right)
    \right], \nonumber \\
W_E(t) &=& t - \frac{\dot T_i^2}{8(1+k^2T_0^2)}\left[
    (4+6k^2T_0^2)t + (2+k^2T_0^2)\sinh\left(\frac{2t}{T_0}\right) \right]. 
\end{eqnarray}
Note that $E(t)$ decreases in contrast with the behavior of magnetic field
and $\dot W_E$ decreases faster than $\dot W_B$.
This can be confirmed in Fig.~\ref{fig1}.

At late time, the value of the 
potential (\ref{vtt}) decays exponentially to zero
\begin{equation}
V(T(t)) \approx \frac{1}{a_+e^{t/T_0}},
\end{equation}
so that Eq.~(\ref{Beq}) becomes, up to ${\cal O}(e^{-2t/T_0})$,
\begin{equation}\label{Beql}
\ddot{B} - \frac{k^2 B_0^4}{\epsilon_0^2} \frac{1}{B^3} = 0,
\end{equation}
The solution of this equation is given by
\begin{equation}
B(t) = \left[\left(\frac{k^2 B_0^4}{\epsilon_0^2 b_0^2}+\dot b_0^2\right)t^2
    +2b_0\dot b_0t+b_0^2 \right]^{1/2} + {\cal O}(e^{-t/T_0}),
\end{equation}
where $b_0$ and $\dot b_0$ are constants.
Therefore $B(t)$ increases linearly in time and 
then the propagation speed decreases
according to Eq.~(\ref{Bspeed}) (See also Fig.~\ref{fig1}).
Since $B$ linearly increases, our linear approximation of the equation
of motion is not valid as $t$ becomes large. However, one may still
extract some relevant physics as the tachyon rolls to the vacuum, 
$V\rightarrow0$. The electromagnetic waves become frozen and stop propagating.
For the electric field in this limit, we find that
it approaches a constant value given by Eq.~(\ref{e-b1}).
Of course, this is an expected behavior since one may interpret
the infinite time limit of the rolling tachyon 
as the Carrollian limit. That is, 
the effective covariant metric tensors
$\bar{\eta}_{\mu\nu}=\mbox{diag}(-1+\dot{T}^{2},1,1,1)$ (\ref{emn})
goes to $\mbox{diag}(0,1,1,1)$, and light cone defined from 
$\bar{\eta}_{\mu\nu}$
collapses down to a timelike half line~\cite{GHY}.

\end{document}